\documentstyle[11pt,newpasp,twoside]{article}
\def\edcomment#1{\iffalse\marginpar{\raggedright\sl#1\/}\else\relax\fi}
\reversemarginpar

\begin{document}

\title{Disc Instabilities in Soft X--ray Transients}
\author{A.R. King} 
\affil{Theoretical Astrophysics Group, University of Leicester, Leicester LE1
7RH, U.K}
            
\begin{abstract}
I briefly review the theory of soft X--ray transient
systems. Irradiation of the accretion disc faces by the central X--ray
source determines both the occurrence and the nature of the outbursts,
in particular forcing these to be long viscous events and producing
exponential decays at short orbital periods. Soft X--ray transients
constitute the majority of LMXBs, persistent systems being largely
confined to a subset of neutron--star LMXBs with periods $P \la
2$~d. It appears that LMXBs very frequently contain nuclear--evolved
companions, even at short orbital periods. In long--period transients
($P \ga 20$~d) the outburst recurrence times must become extremely
long ($\ga 1500$~yr). The outbursts are highly super--Eddington,
markedly reducing the accretion efficiency. Spinup of a neutron--star
primary to millisecond periods probably cannot occur for orbital
periods $\ga 200$~d, in agreement with observations of binary pulsars.

\end{abstract}

\section{Introduction}

My interest in this subject was kindled by a workshop on Black Holes
held in Aspen in early 1996, and transformed by interaction with
Jan. I arrived at the meeting after a long snowy drive from Denver,
worried about two things. First, my then postdoc Luciano Burderi had
persuaded me to try skiing for the first time, at what even I regarded
as a perilously advanced age. Second, the meeting organisers had asked
me to talk about the evolution of X--ray transients, and I could not
make much sense out of this topic. By then the only model seriously
considered for the X--ray outbursts involved the thermal--viscous disc
instability, resulting from hydrogen ionization. I was puzzled by two
aspects of this. I could not see why the outbursts of transients
should be so much longer and more infrequent than those of dwarf
novae, and I could not understand why some low--mass X--ray binaries
(LMXBs) should be transient and others not.

Being asked to talk on the subject made the second of these problems
more immediately urgent than the first. My then postdoc Uli Kolb and I
had tried to solve it by simply asking when an LMXB accretion disc
would have hydrogen ionization zones. The presence of these zones
should mean that the disc was unstable, making the system transient.
This procedure already worked well for the closely related cataclysmic
variables (CVs), in which the accretor is a white dwarf rather than
the black hole or neutron star in LMXBs. Given the disc surface
temperature distribution $T_{\rm eff}(R)$ resulting from viscous
dissipation, all one had to do was compare this with some value
$T_{\rm H} \simeq 6500$~K typical of hydrogen ionization. If $T_{\rm
eff}(R)$ was everywhere above $T_{\rm H}$ the disc should have no
ionization zones and be persistent rather than having dwarf nova
outbursts. Since $T_{\rm eff}(R)$ decreases outwards the condition
amounted to 
\begin{equation}
T_{\rm eff}(R_{\rm d}) > T_{\rm H} 
\label{eff}
\end{equation}
where $R_{\rm d}$ is the outer disc radius. As $T_{\rm eff}(R_{\rm
d})$ could be calculated from the mass transfer rate $-\dot M_2$ and
orbital period $P$, one could plot a line dividing dwarf novae from
persistent (`novalike') CVs on the usual $-\dot M_2 - P$ relation
predicted for CVs, assuming the standard angular momentum loss
mechanisms and seondary stars close to the main sequence. Those CVs
predicted to have ionization zones, i.e. lie below the line, were
indeed dwarf novae. Yet for LMXBs the same method produced the
manifestly wrong result that {\it all} short--period ($P \la 12$~hr)
LMXBs should be transient.

Surviving the first couple of days on the ski slopes without serious
injury eased my first set of worries, but I was still fretting over
the second set when Jan gave his talk. By immense good fortune this
was the evening before I was due to speak. Jan's talk
characteristically combined a powerful insight with a presentation so
lucid and simple that the result seemed instantly obvious. He pointed
out that LMXB discs differ from CV discs in one vital respect: the
optical flux from an LMXB disc is so much higher than expected from
the accretion rate revealed by the X--ray flux, that the disc must be
heated by some other agency than the viscous dissipation driving the
accretion through the disc. The obvious candidate for this is the
X--ray emission itself, some of which must fall on the disc faces and
heat them. This fact had long been known by observers, but somehow
never fully appreciated by theorists, largely I suspect because
theoretical calculations predicted (and still predict) that the result
of irradiating a disc in this way is to make its central regions swell
up and shield most of the rest of it from the X--rays.

Armed with clear observational evidence to the contrary, Jan adopted
the sensible view that nature knows how to irradiate a disc even if
theorists don't. A simple concave disc model allowed him to calculate
the run of disc surface temperature $T_{\rm irr}(R)$ from the observed
X--ray flux and orbital period of a given LMXB. If $T_{\rm irr}(R)$
was everywhere above $T_{\rm H}$, Jan assumed that the disc would be
stable and the system persistent. Since $T_{\rm irr}(R)$ decreases
outwards this is now equivalent to requiring
\begin{equation}
T_{\rm irr}(R_{\rm d}) > T_{\rm H} \label{irr}
\end{equation}
rather than the CV condition (\ref{eff}). Jan showed that indeed the
condition (\ref{irr}) correctly divided LMXBs into persistent and
transient if he used their {\it observed} X--ray luminosities to predict
$-\dot M_2$ and thus $T_{\rm irr}$. 

Back in my room that evening I spent a lot of time thinking about what
Jan had said, and how I should change the talk I was to deliver the
next day. It was clear that condition (\ref{irr}) was less restrictive
than (\ref{eff}), reflecting the stabilizing effect of irradiation. So
the question was what the corresponding line on the $-\dot M_2 - P$
plane would give. Fortunately Jan had given enough information for me
to add the line to the plots I had prepared for my talk. Sure enough,
the new line was at lower $-\dot M_2$ -- values than the old line. But
the surprising consequence was that, particularly for neutron--star
LMXBs, the standard angular--momentum driven evolution of
short--period systems with near--main--sequence secondaries now
predicted that all systems in the observed 3~hr $\la P \la$ 10~hr
period range should be be {\it persistent}. While this was an advance
on the clearly incorrect earlier prediction that all these systems
should be transient, I still had to explain the presence of several
incontrovertible neutron--star transients at these periods. Given the
robust nature of Jan's condition (\ref{irr}), the only plausible route
seemed to lie in dropping one of the assumptions about the evolution
of neutron--star LMXBs. By another stroke of good luck, I had brought
with me an $-\dot M_2 - P$ plot for an LMXB where the secondary star
was somewhat nuclear--evolved at the start of mass transfer, but the
system evolved to shorter periods under angular momentum loss. I
noticed that the predicted mass transfer rates $\dot M_2$ were
significantly lower; this occurred since the somewhat evolved
secondary star was slightly larger for its mass than a main--sequence
secondary would be. Adding the line corresponding to condition
(\ref{irr}) immediately showed that this system would indeed be
transient at short periods. In other words, the existence of
short--period neutron--star transients is evidence that nuclear
evolution can have a significant effect even at such periods. 

Jan's insight (published as van Paradijs, 1996) thus started off an
important line of research, which I briefly review in the rest of this
article.

\section{The Current Situation}

\subsection{Long--period transients and millisecond pulsars}

The consequences of the condition (\ref{irr}) for the long--term
evolution of LMXBs have been explored in a number of papers. King,
Kolb \& Burderi (1996) reinforced the conclusion above that evolved
secondaries favour transient behaviour, and noted that all
long--period ($P \ga 2$~d) LMXBs are likely to be transient because
the accretion disc is very large and must have cool edges. Such
systems must be the progenitors of long--period ($P \ga 100 - 200$~d)
millisecond pulsar binaries with circular orbits (e.g. Tauris \&
Savonije, 1999), yet the longest orbital period seen in a
neutron--star LMXB is only 11.8~d (GRO~J1744-28, Giles et al., 1996,
which is indeed transient). This must mean that outbursts in these
long--period transients are so infrequent none has been seen over the
$\sim 30$~year history of X--ray astronomy. Moreover, these outbursts
are likely to be highly super--Eddington (even the mean mass transfer
rates are), so very little of the transferred mass will be accreted by
the neutron star. As a consequence it is very hard to spin up these
stars to millisecond periods, particularly at long orbital periods,
explaining an observed trend. In addition, by exploiting these
observational constraints, Ritter \& King (2001a,b) have shown that
the mean recurrence time of outbursts in long--period transients must
become extremely long; the current lower limit is about 1500~yr. This
is an interesting challenge for accretion disc theory.

\subsection{The evolution of short--period transients}

Clearly the conclusion that nuclear evolution is significant even in
short--period systems implies interesting constraints on the
pre--contact evolution. King \& Kolb (1997) and Kalogera, Kolb \& King
(1998) showed that the existence of a sufficient number of such
systems to account for short--period neutron--star transients is
plausible, even without extreme assumptions: the formation of a
neutron--star binary is such a rare event that rather unusual
companions are not uncommon. King, Kolb \& Szuszkiewicz (1997) noted
that the formation constraints for black--hole binaries were
considerably weaker, chiefly because any supernova explosion in the
latter does not come perilously close to unbinding the binary, as is
inevitable in neutron--star systems. With standard assumptions about
common--envelope evolution (CE) and magnetic braking (MB) there is
little to prevent the formation of large numbers of black--hole
binaries with unevolved low--mass companions. These systems would be
persistent, in conflict with the observation that most persistent
short--period LMXBs show Type I X--ray bursts and therefore contain
neutron stars. Without modifying CE and MB, the most likely escape
seems to be an effect noted by Shakura \& Sunyaev (1973) in their
original paper on accretion discs. This involves the lack of a hard
surface for black holes: this may inhibit the formation of a central
point irradiating source. The accretion disc will still be irradiated,
but now only by the inner accretion disc rather than a
quasi--spherical object, weakening the irradiation effect by the
aspect ratio $H/R$ of the disc ($H = $ disc scaleheight). Taking
account of this factor in the irradiation formula does indeed make
such BH + main sequence binaries transient. However recently there has
emerged evidence that many if not most short--period black--hole
transients have significantly nuclear--evolved companions: HST spectra
of XTE J1118+480 show clear signs of CNO processing for example. This
in turn must mean that the formation constraints for LMXBs differ from
those resulting from the standard assumptions about MB and
particularly CE. This may remove the motivation for including the
$H/R$ factor in the irradiation formula. The conclusion about CE may
have important consequences for general close binary evolution,
including CVs (King \& Schenker, 2001). In particular nuclear
evolution and finite age effects may be far more significant that
hitherto thought.

\subsection{Soft X--ray transient outbursts}

None of the work referred to above deals with the crucial question
raised in the Introduction of why transient outbursts have much longer
timescales than dwarf novae. Here too disc irradiation provides an
insight. Once an outburst starts, the central regions of the disc will
find it impossible to return to the cool state until the central
X--ray source turns off. But this turnoff itself cannot occur while
these central regions contain significant mass: observation is
unambiguous in showing that the disc remains strongly irradiated
throughout the outburst. Clearly the outburst must last until most of
this mass is removed by accretion, which occurs on the hot--state
viscous timescale. King \& Ritter (1998) constructed a simple
analytical model of this process, incorporating the assumption that
the disc was irradiated by the central source in the way indicated by
the earlier evolutionary studies and Jan's original paper. King \&
Ritter's paper showed that in short--period systems, where the whole
of the disc faces could be effectively illuminated, there was a strong
tendency to produce an exponential X--ray decay, while in longer
period systems only the centre of the disc could be kept in the hot
state, producing a linear decay. This is supported by observation
(Giles et al., 1996; Shahbaz et al., 1998). More elaborate numerical
calculations confirm this (cf Lasota, 2001, and references therein),
although the theoretical problem of why the central disc does not puff
up and shield the outer parts from the irradiation is still not
understood. Recently, Truss et al. (2001) have suggested that the
secondary maximum seen in exponential decays may result from the
standard unirradiated ionization instability operating on matter at
the edge of a small disc, which is shielded from the central X--rays
simply because there is inevitably a shadowing effect at the outer
disc edge.

\section{Conclusions}

Jan's insight that disc irradiation could have important effects has
been amply justified. He would have been pleased by this, but even
happier to realise that we are only at the beginning of understanding
all its consequences.

\section{Acknowledgment}
I thank the conference organisers and all the participants for what
was a memorable experience. As Jan would have wished, the meeting was
a marvellously stimulating scientific occasion.

\end{document}